\documentclass[pra,showpacs,twocolumn]{revtex4}
\usepackage{amsfonts}
\usepackage{amsmath}
\usepackage{amssymb}
\usepackage{graphicx}
\renewcommand{\Re}{\mathop{\mathrm{Re}}}
\renewcommand{\Im}{\mathop{\mathrm{Im}}}

\begin{document}
\title{Quantum-limited metrology in the presence of collisional dephasing}

\author{Y. C. Liu}
\author{G. R. Jin}
\email{grjin@bjtu.edu.cn} \affiliation{Department of Physics,
Beijing Jiaotong University, Beijing 100044, China}
\author{L. You}
\affiliation{Department of Physics, Tsinghua University, Beijing
100084, China}

\begin{abstract}
Including collisional decoherence explicitly, phase sensitivity for
estimating effective scattering strength $\chi$ of a two-component
Bose-Einstein condensate is derived analytically. With a measurement
of spin operator $\hat{J}_{x}$, we find that the optimal sensitivity
depends on initial coherent spin state. It degrades by a factor of
$(2\gamma)^{1/3}$ below super-Heisenberg limit $\propto 1/N^{3/2}$
for particle number $N$ and the dephasing rate
$1<\!<\gamma<N^{3/4}$. With a $\hat{J}_y$ measurement, our
analytical results confirm that the phase $\phi=\chi t\sim 0$ can be
detected at the limit even in the presence of the dephasing.
\end{abstract}
\date{\today }
\pacs{03.75.Dg, 03.75.Mn, 03.75.Gg} \maketitle


Parameter estimation with its precision beyond standard quantum
limit (SQL) is a long-standing challenge in quantum metrology. The
achievable precision depends on the input state
\cite{Caves,Yurke,Holland,Wineland1,Wineland2,Mitchell,Giovannetti},
the observable being measured at output ports
\cite{Bollinger,Campos,Dowling}, and the coupling nature of the
system Hamiltonian \cite{Luis,Rey,Boixo08,Choi,Woolley}. In standard
Ramsey interferometry, for instance, resonant atomic frequency
$\chi$ is estimated with a coherent spin state (CSS) that evolves
freely under a linear coupling $\chi\hat{J}_z$. It has been shown
that the precision scales as $1/N^{1/2}$ (i.e., the SQL) for
$\hat{J}_x$ or $\hat{J}_y$ measurement \cite{Boixo08}. The phase
sensitivity can be improved to the so-called Heisenberg limit $1/N$
with an entangled input state
\cite{Wineland2,Mitchell,Giovannetti,Bollinger}. Here, collective
spin operators
$\hat{J}_{v}=\frac{1}{2}\sum_{k}\hat{\sigma}_{v}^{(k)}$ with the
Pauli matrices $\hat{\sigma}_{v=x,y,z}$.

Recently, atom interferometry with Bose-Einstein condensates (BEC)
becomes a topical area of study due to potential applications in
quantum metrology \cite{Giovannetti} and quantum information
\cite{Sorensen}. Elastic collision of condensed atoms can be
described by the `one-axis twisting' Hamiltonian
$\chi\hat{J}_{z}^{2}$ \cite{Kitagawa}, capable of creating
multipartite entanglement and spin squeezing
\cite{Sorensen,Kitagawa,Jin09,Gross,Riedel}. Control of spin
dynamics requires precise measurement of the effective interaction
strength $\chi$ \cite{Artur0}. Rey \textit{et al}. \cite{Rey}
recently claimed that even with an initially CSS, the Heisenberg
limit is achievable if the free evolution under $\chi\hat{J}_{z}$ is
simply replaced by $\chi\hat{J}_{z}^{2}$. A better scaling $\propto
1/N^{3/2}$ is proposed for the CSS prepared by a $\pi/4$ or $3\pi/4$
pulse \cite{Boixo08,Choi}. Such a super-Heisenberg scaling can also
reach in nonlinear optical and nano-mechanical systems
\cite{Luis,Woolley}.

In this brief report, we investigate carefully the effects of
collisional dephasing on the BEC-based quantum metrology
\cite{Rey,Boixo08,Choi}, described by the Hamiltonian ($\hbar=1$):
$\hat{H}=\chi\hat{J}_{z}^{2}+\Omega_{x}\hat{J}_{x}+\Omega
_{y}\hat{J}_{y}$, where the effective interaction strength $\chi$
and the Rabi frequencies $\Omega_{x}=\Re(\Omega)$ and
$\Omega_{y}=\Im(\Omega)$ are tunable in real experiments
\cite{Gross,Riedel}. The metrology protocol starts with a Ramsey
pulse applied to the BEC with all spin up, yielding a product CSS:
$|\Psi_{\theta}\rangle=e^{-i\theta\hat{J}_{y}}|J,J\rangle=|\theta,
0\rangle$, where the polar angle $\theta=\Omega_{y}t$ given by the
pulse area. Next, the system evolves freely for a time $t$,
$|\Psi_{\theta}(\phi)\rangle=e^{-i\phi\hat{J}_{z}^{2}}|\Psi_{\theta}\rangle$,
with a dimensionless phase shift $\phi=\chi t$. Finally, an
equatorial component of the total spin $\hat{J}_{x}$ or
$\hat{J}_{y}$ is measured to estimate $\chi$ \cite{Boixo08}.

For nonzero $\chi$, the accumulated phase $\phi$ manifests itself as
oscillations of the Ramsey signal $\langle\hat{J}_x\rangle$ or
$\langle\hat{J}_y\rangle$, with its sensitivity quantified by
\cite{Boixo08}
\begin{eqnarray}
\delta\phi_{v}=t\delta\chi_{v}=\frac{\Delta\hat{J}_{v}}{|d\langle
\hat{J}_{v}\rangle/d\phi|}, \hskip 5pt (v=x, \text{or}\ y),
\label{Sensitivity}
\end{eqnarray}%
where the variance
$\Delta\hat{A}\equiv(\langle\hat{A}^2\rangle-\langle\hat{A}\rangle^2)^{1/2}$
and the expectation value $\langle\hat{A}
\rangle\equiv\langle\Psi_{\theta}(\phi)|\hat{A}|\Psi_{\theta}(\phi)\rangle$.
The $\hat{J}_{x/y}$ measurement is achievable by applying a second
$\pi/2$ pulse after the free evolution,
$e^{-i\frac{\pi}{2}\hat{J}_{y/x}}|\Psi_{\theta}(\phi)\rangle$, and
following with a detection of population imbalance (i.e.,
$\hat{J}_{z}$), as done in standard Ramsey interferometry.


Collective spin excitation and external field fluctuations leads to
an enhanced phase diffusion of the BEC \cite{Artur}. To describe it
qualitatively, we assume that free evolution of the system obeys the
master equation \cite{Puri,Vardi}: $\dot{\rho}=i[\rho,\chi
\hat{J}_{z}^{2}]+\Gamma (2\hat{J}_{z}
\rho\hat{J}_{z}-\hat{J}_{z}^{2}\rho-\rho\hat{J} _{z}^{2})$, where
$\Gamma$ denotes the dephasing rate. For single-particle case, the
second term reduces to $\frac{\Gamma
}{2}(\hat{\sigma}_{z}\rho\hat{\sigma}_{z}-\rho)$ \cite{Huelga}. The
many-body decoherence considered here, known as the collisional
dephasing \cite{Puri}, can be solved exactly with the density matrix
element $\rho_{m, n}(\phi)\equiv\langle J,m|\rho|J,n\rangle=\rho_{m,
n}(0)\exp[i(n^{2}-m^{2})\phi-\gamma (m-n)^{2}\phi]$, which yields
\begin{eqnarray}
\langle\hat{J}_{+}\rangle&=&J e^{-\gamma\phi}\sin\theta
(\cos\phi+i\cos\theta\sin\phi)^{2J-1},  \label{J+} \\
\langle\hat{J}_{+}^{2}\rangle&=&J(J-1/2)e^{-4\gamma\phi
}\sin^{2}\theta \notag \\
&&\times(\cos2\phi+i\cos\theta\sin 2\phi)^{2J-2},  \label{J+2}
\end{eqnarray}%
where $J=N/2$, $\gamma=\Gamma/\chi$, and $\phi=\chi t$. Note that
the collisional dephasing imposes an exponential decay to average
value of $\hat{J}_{+}$ and its higher moment $\hat{J}_{+}^2$, but
maintains that of $\hat{J}_{z}$ and $\hat{J}_{z}^{2}$ intact, i.e.,
$\langle\hat{J}_{z}\rangle=J\cos\theta$ and
$\langle\hat{J}_{z}^{2}\rangle=J^{2}-J(J-1/2)\sin^{2}\theta$ (c.f.
Ref.~\cite{Jin09}). The slope of the signal $d\langle
\hat{J}_{x/y}\rangle/d\phi$ corresponds to real or imaginary part of
$d\langle\hat{J}_{+}\rangle/d\phi$, with
\begin{eqnarray}
\frac{d\langle \hat{J}_{+}\rangle}{d\phi} &=&J(2J-1)e^{-\gamma
\phi }(i\cos\theta\cos\phi-\sin\phi)  \notag \\
&&\times\sin(\theta)(\cos\phi+i\cos\theta\sin\phi)^{2J-2}.
\label{J+2Jzp1}
\end{eqnarray}%
We exclude the case $\theta=0$ and $\pi$, due to $d\langle
\hat{J}_{v}\rangle_{\gamma}/d\phi=0$ and thus $\delta
\phi_{v}\rightarrow\infty$, which implies no information about
$\phi$ (or $\chi$ for a given $t$) is gained from $\hat{J}_{x}$ and
$\hat{J}_{y}$ measurements.

\begin{figure}[t]
\begin{center}
\includegraphics[width=3in]{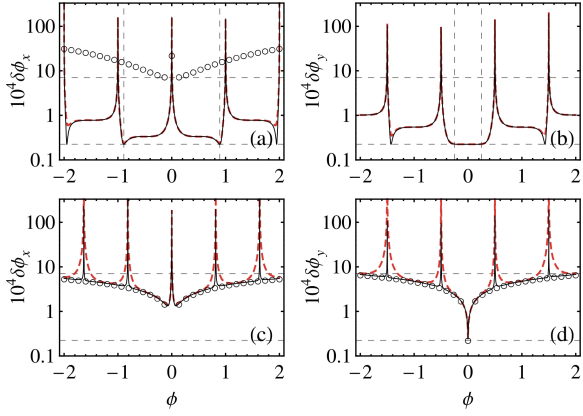}
\caption{(Color online) Phase sensitivities $\delta\phi_{x}$ (left)
and $\delta\phi_{y}$ (right) as a function of phase shift $\phi$ [in
units of $\pi/(\sqrt{2}J)$] for $J=N/2=10^{3}$ and $\gamma=0$ (up),
$10^{2}$ (bottom). Solid (dashed red) lines are analytical (exact
numerical) results for $\theta=\pi/4$ (\textbf{a}, \textbf{b},
\textbf{d}) and $\theta=\pi/6$ (\textbf{c}). Empty circles in
(\textbf{a}) is exact result for $\theta=\pi/2$, while in
(\textbf{c}) and (\textbf{d}) are obtained from
Eq.~(\ref{sensGammaJx}) and Eq.~(\ref{sensGammaJy}) for
$\theta=\pi/6$ and $\pi/4$, respectively. Horizontal lines denote
the Heisenberg limit $1/(\sqrt{2}J)$ \cite{Rey}, and the
super-Heisenberg limit $1/(\sqrt{2}J^{3/2})$ \cite{Boixo08}.
Vertical lines in (\textbf{a}) and (\textbf{b}) correspond to
$|\phi|=0.89\pi/(\sqrt{2}J)$ and $0.25\pi/(\sqrt{2}J)$,
respectively.} \label{fig1}
\end{center}
\end{figure}

To obtain scaling rule of $\delta\phi_{v}$, we perform standard
short-time analysis to the above exact results
\cite{Kitagawa,Jin09}. After some straightforward calculations, we
get $(\Delta\hat{J}_{x})^{2}\simeq \frac{J}{2}[1-(\eta
_{0}-J\eta_{1})\sin^{2}\theta]$, where
$\eta_{0}=\frac{1}{2}(1+e^{-4\beta}\cos2\alpha)$ and $\eta
_{1}=(1-e^{-2\beta})(1-e^{-2\beta}\cos2\alpha)+2\phi
e^{-2\beta}\cos(\theta)\sin2\alpha$. In the short-time limit
($|\phi|<<1$), the parameters $\alpha=2J\phi\cos\theta\sim 1$ and
$\beta=J\phi^{2}\sin^{2}\theta+\gamma\phi<<1$, which enable us to
expand $\eta_{0}$ ($\eta_{1}$) up to the zeroth (the first)-order of
$\beta$. Similarly, Eq.~(\ref{J+2Jzp1}) gives
$d\langle\hat{J}_{x}\rangle/d\phi\simeq
-J^{2}\sin(\alpha)\sin2\theta$ for $\theta\neq\pi/2$. As a result,
we obtain
\begin{equation}
\delta\phi_{x}^{2}\simeq\frac{1+(\cos\theta\cot\alpha +2J\phi
\sin^{2}\!\theta)^{2}+4\gamma J\phi\sin^{2}\!\theta}{2J^{3}\sin
^{2}2\theta}. \label{Phix}
\end{equation}%
Replacing $\alpha$ with $\alpha +\pi/2$, we also get the short-time
solution of $(\Delta\hat{J}_{y})^{2}$ and that of the sensitivity
\begin{equation}
\delta\phi_{y}^{2}\simeq\frac{1+(\cos\theta\tan\alpha-2J\phi\sin
^{2}\!\theta)^{2}+4\gamma
J\phi\sin^{2}\!\theta}{2J^{3}\sin^{2}2\theta}.  \label{Phiy}
\end{equation}%
As depicted in Fig.~\ref{fig1}, the sensitivities oscillate rapidly
in a fringe pattern and diverge at $|\phi|=s\pi/(2J\cos\theta)$ and
$(s+1/{2})\pi/(2J\cos\theta)$ for an integer $s=0$, $1$, etc., given
by $\cot\alpha\rightarrow\infty$ and $\tan\alpha\rightarrow\infty$
in Eq.~(\ref{Phix}) and Eq.~(\ref{Phiy}), respectively. Within
central few fringes ($s\leq 2$), our analytical results (thin solid
lines) are coincident with the exact numerical simulations (red
dashed lines).

\begin{figure}[hp]
\begin{center}
\includegraphics[width=3in]{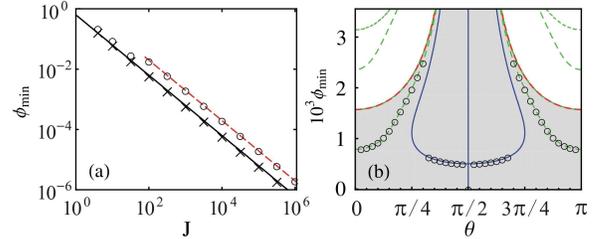}
\caption{(Color online) The best-sensitivity phase $\phi_{\min}$ as
a function of $J$ (\textbf{a}) and $\theta$ (\textbf{b}) for
$\hat{J}_x$ measurement and $\gamma=0$. In (\textbf{a}), numerical
results of $\phi_{\min}$ for $\theta=\pi/4$ (empty circles) and
$\pi/3$ (crosses), fit with $0.89\pi/(\sqrt{2}J)$ (dashed red line)
and $0.2\pi/J$ (solid line). In (\textbf{b}), contour plots of the
first (dashed green line) and the second (solid blue line) equations
of (\ref{timeJx}) for $J=10^3$, with numerical $\phi_{\min}$ (empty
circles). The shade region denotes the central fringe
$0\leq\phi_{\min}<\pi/(2J\cos\theta)$.} \label{fig2}
\end{center}
\end{figure}

We now analyze the achievable sensitivity for $\hat{J}_x$ and
$\hat{J}_y$ measurements one by one. Firstly, let us consider the
case $\gamma=0$. Via minimizing Eq.~(\ref{Phix}) with respect to
$\phi$, we find that local minima of $\delta\phi_{x}$ occur when
\begin{equation}
\cos(\theta)\cot\alpha+2J\phi\sin^{2}\theta =0\text{,
}\hskip3pt\text{ or }\sin\alpha=\cot\theta.  \label{timeJx}
\end{equation}
The first transcendental equation gives
$(\delta\phi_{x})_{\min}\simeq {1}/({\sqrt{2}J^{3/2}|\sin
2\theta|})$, as predicted in Ref.~\cite{Boixo08}. For $\theta=\pi/4$
or $3\pi/4$, it becomes $1/({\sqrt{2}J^{3/2}})$. Such a
super-Heisenberg scaling of the sensitivity appears when
$\alpha=\sqrt{2}J\phi\simeq\pm 0.89\pi$, i.e.,
$\phi_{\min}=|\chi|_{\min}t\simeq 0.89\pi/(\sqrt{2}J)$ [see
Fig.~\ref{fig1}(a)], which is valid provided $J>10^2$
[Fig.~\ref{fig2}(a)]. Numerically, we also find that the best
sensitivity phase $\phi_{\min}$ depends on polar angle of the
initial state $\theta$ [empty circles of Fig.~\ref{fig2}(b)]. For
$\pi/4<\theta<3\pi/4$, it is in fact determined by
$\sin\alpha=\cot\theta$, which predicts $\phi_{\min}\simeq 0.2\pi/J$
for $\theta=\pi/3$ [crosses of Fig.~\ref{fig2}(a)].

To proceed, we consider the $\hat{J}_y$ measurement in the absence
of the dephasing. From Eq.~(\ref{Phiy}), one can find that minimal
value of the sensitivity $(\delta\phi_{y})_{\min}\simeq
{1}/({\sqrt{2}J^{3/2}|\sin 2\theta|})$ occurs at $\phi\sim 0$.
Obviously, the super-Heisenberg limit $1/({\sqrt{2}J^{3/2}})$ is
attainable for the optimal CSS with $\theta=\pi/4$ or $3\pi/4$ [see
Fig.~\ref{fig1}(b), also Ref.~\cite{Boixo08}).

\begin{figure}[t]
\begin{center}
\includegraphics[width=2.5in]{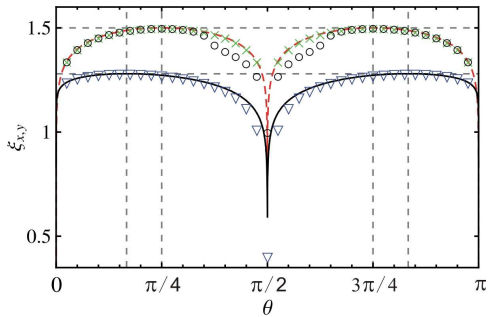}
\caption{(Color online) Scaling exponents $\xi_x$ and $\xi_y$ as a
function of polar angle $\theta$ for $J=10^5$. The circles (the
crosses) are numerical results of $\xi_x$ ($\xi_y$) for $\gamma=0$,
and the dashed red line is obtained from Eq.~(\ref{xi}). The
triangles are numerical results of $\xi_x$ for $\gamma=10^3$, and
the solid line is predicted by Eq.~(\ref{xiGamma}). Vertical lines
denote $\theta=\pi/6$, $\pi/4$, $3\pi/4$, and $5\pi/6$; horizontal
ones correspond to $3/2$, $3/2-\ln(2\gamma)/(3\ln J)$.} \label{fig3}
\end{center}
\end{figure}

The above results can be casted in a more transparent form by
setting $(\delta\phi_{v})_{\min}=\kappa J^{-\xi_{v}}$ (for $v=x$,
$y$) \cite{Boixo08}, with a pre-factor $\kappa$ and scaling exponent
\begin{equation}
\xi_{v}=-\frac{\ln[\sqrt{2}(\delta \phi_{v})_{\min}]}{\ln J}\simeq
\frac{3}{2}+\frac{\ln\left(\left\vert\sin\!2\theta\right\vert
\right)}{\ln J}, \label{xi}
\end{equation}
where we set $\kappa=1/\sqrt{2}$ to ensure $\xi_{v}\rightarrow 3/2$
as $\theta\rightarrow\pi/4$ or $3\pi/4$. One can find from
Fig.~\ref{fig3} that without the dephasing, numerical results of
$\xi_{x}$ (empty circles) and $\xi_{y}$ (green crosses) agree with
Eq.~(\ref{xi}) (dashed red line).


Finally, we discuss the achievable sensitivity in the presence of
the dephasing. For $\theta\neq\pi/2$ and $\gamma>>1$,
Eq.~(\ref{Phix}) and Eq.~(\ref{Phiy}) can be simplified as
\begin{eqnarray}
\delta\phi_{x}^{2}&\simeq &\frac{1}{2J^{3}\sin^{2}2\theta
}\left[(4J\phi)^{-2}+4\gamma J\phi\sin^{2}\theta\right], \label{sensGammaJx}\\
\delta\phi_{y}^{2}&\simeq &\frac{1}{2J^{3}\sin^{2}2\theta
}\left[1+4\gamma J\phi\sin^{2}\theta\right], \label{sensGammaJy}
\end{eqnarray}
which correspond to the envelope curves of the sensitivities [see
empty circles of Figs.~\ref{fig1}(c) and (d)]. For $\hat{J}_x$
measurement, the best sensitivity $(\delta\phi_{x})_{\min}^{2}\simeq
{3\gamma^{2/3}}/({8J^{3}\sin^{2/3}\theta\cos^{2}\theta})$, and thus
\begin{equation}
\xi_{x}\simeq\frac{3}{2}+\frac{\ln(\frac{2}{\sqrt{3}}\gamma^{-1/3}\sin^{1/3}\theta\left\vert
\cos\theta\right\vert)}{\ln J}.\label{xiGamma}
\end{equation}
As shown in Fig.~\ref{fig3}, our analytical result shows a good
agreement with numerical simulations (triangles) for the dephasing
rate $1<\!<\gamma<J^{3/4}$. Remarkably, the optimal scaling
$\xi_{x}\simeq 3/2-\ln(2\gamma)/(3\ln J)$ is obtained at
$\theta=\pi/6$ or $5\pi/6$, which leads to the best-sensitivity
phase $\phi_{\min}\simeq 1/[J(2\gamma)^{1/3}]$ with
$(\delta\phi_{x})_{\min}\simeq(2\gamma)^{1/3}/(\sqrt{2}J^{3/2})$.
For $\hat{J}_y$ measurement, the phase $|\phi|\sim s\pi/(\sqrt{2}J)$
(with an integer $s\leq 2$) can be detected at the super-Heisenberg
limit provided $\theta=\pi/4$ or $3\pi/4$. However, the sensitivity
for $|\phi|>2\pi/(\sqrt{2}J)$ degrades rapidly [see
Fig.~\ref{fig1}(d)].

In summary, we have derived analytical results for precise
estimation of effective interaction strength $\chi$ in a
two-component BEC. Without collisional dephasing, the
best-sensitivity phase $\phi_{\min}$ for $\hat{J}_x$ measurement
depends on the initial CSS $|\theta, 0\rangle$ and bifurcates at
$\theta\sim\pi/4$ or $3\pi/4$ (see Fig.~\ref{fig2}b). In the
presence of the dephasing (with $1\leq\gamma<J^{3/4}$), the optimal
CSS becomes $|\theta=\pi/6, 0\rangle$, and the achievable
sensitivity is reduced by a factor of $(2\gamma)^{1/3}$ below the
scaling $1/(\sqrt{2}J^{3/2})$. Our analytical results confirm that
the detection of $\hat{J}_y$, i.e., the $\hat{J}_z$ measurement to
the output state
$e^{-i\frac{\pi}{2}\hat{J}_{x}}e^{-i\phi\hat{J}_{z}^{2}}e^{-i\frac{\pi}{4}\hat{J}_{y}}|J,J\rangle$,
shows its advantages since phase estimation of $\phi\sim0$ can reach
the super-Heisenberg limit even in the presence of the dephasing
\cite{Boixo08}.

This work is supported by the NSFC (Contract No.~10804007) and the
SRFDP (Contract No.~200800041003). L.Y. is partially supported by
the NKBRSF of China under Grants 2006CB921206 and 2006AA06Z104.


\end{document}